\documentstyle[11pt]{article}
\def\thesection{\Roman{section}}
\newcommand{\R}{\rm I\hspace{-0.4ex}R}
\newcommand{\beq}{\begin{equation}}
\newcommand{\eeq}{\end{equation}}
\newcommand{\beqa}{\begin{eqnarray}}
\newcommand{\eeqa}{\end{eqnarray}}
\newcommand{\del}{\frac{\partial\phi}{\partial x}}
\newcommand{\delt}{\frac{\partial\phi}{\partial t}}
\newcommand{\dels}{\frac{\partial\phi_{s}}{\partial s}}
\newcommand{\delx}{\frac{\partial\phi_{s}}{\partial x}}
\newcommand{\grad}{{\rm grad}\,E[\phi]}
\newcommand{\sgrad}{{\rm grad}}
\newcommand{\inta}{\int_{\Sigma}dt\,dx}
\newcommand{\ii}{{\cal I}}
\newcommand{\jj}{{\cal J}}
\renewcommand{\theequation}{\mbox{\arabic{section}.\arabic{equation}}}
\makeatletter
\def\eqnarray{%
\stepcounter{equation}%
\let\@currentlabel=\theequation
\global\@eqnswtrue
\global\@eqcnt\z@
\tabskip\@centering
\let\\=\@eqncr
$$\halign to \displaywidth\bgroup\@eqnsel\hskip\@centering
$\displaystyle\tabskip\z@{##}$&\global\@eqcnt\@ne
\hfil$\displaystyle{{}##{}}$\hfil
&\global\@eqcnt\tw@$\displaystyle\tabskip\z@{##}$\hfil
\tabskip\@centering&\llap{##}\tabskip\z@\cr}
\makeatother
\begin{document}
\title{Unstable geodesics and topological field theory}
\author{Yukinori Yasui\thanks{E-mail address:
f51999@sakura.kudpc.kyoto-u.ac.jp}\\{\it Depertment of Physics,
Osaka City University,}\\{\it Sumiyoshiku, Osaka, Japan}\\\\
Satoshi Takahashi\thanks{E-mail address: e50874@JPNKUDPC.BITNET}
\\{\it Depertment of Mathematics,
Osaka City University,}\\{\it Sumiyoshiku, Osaka, Japan}}
\date{December 1993}
\maketitle
%\newpage
\begin{abstract}
A topological field theory is used to study the cohomology
of mapping space. The cohomology is identified with the
BRST cohomology realizing the physical Hilbert space
and the coboundary operator given by the calculations of
tunneling between the perturbative vacua. Our method is
illustrated by a simple example.
\end{abstract}

\newpage
%p1
%
\section{INTRODUCTION}
\setcounter{equation} {0}
Recently much attension has been paid on the
study of topological quantum
field theories.$^{{\rm 1,2}}$ In this paper, we discuss
a topological field theory
over 2-dimensional space $\Sigma = \R\times[0,1]$ and
give a field theoretic realization of the cohomology of the mapping
space $\Omega_{p,q}(N) = \{ \phi:[0,1]\longrightarrow N
\mid \phi(0) = p, \phi(1) = q \}$ for a Riemannian manifold $(N,g)$,
which arises as the BRST cohomology representing
 the physical Hilbert space.

The techniques we employ to evaluate the BRST cohomology
are a mixture of Morse theory adapted to
$\Omega_{p,q}(N)$ and instanton calculations
 --- calculations of tunneling
between the perturbative vacua of the topological theory.
This approach was initiated by Witten in finite-dimensional
situations.$^{{\rm 3}}$
Using a supersymmetric quantum mechanics, he introduced
a new cohomology complex and provided a path integral
method of computing
the cohomology of a finite-dimensional manifold $M$.

Let us summarize Witten's idea. Suppose we are given
the following data
concerning Morse theory:
\begin{enumerate}
\item[(M-1)] a Morse function $E$ on $M$,
%p2
%
\item[(M-2)] the critical points of $E$,
\item[(M-3)] the Morse index of each critical point, i.e.,
the number of negative eigenvalues of the Hessian of $E$ evaluated
at the critical point.
\end{enumerate}

Once we have these data, we can construct the
 perturbative vacuum for
each critical point. The vacua are represented
as forms on $M$, whose
degrees are equal to the Morse indices of the
corresponding critical points.
Then we define, for each degree $\ell$, the vector space $X^{\ell}$
consisting of $\ell$-form vacua. Under fortunate circumstances (for example
there are no vector spaces of odd indices), each vacuum becomes a generator
of the cohomology of $M$. However, we must in general consider a coboundary
operator $\delta$ : $X^{\ell} \longrightarrow X^{\ell+1}$ such that
$\delta \circ \delta$ = 0 and
\beq
H^{*}(M,\R) = Ker\ \delta/Im\ \delta.
 \eeq
This operator is interpreted as tunneling effects in a supersymmetric
quantum mechanics and is evaluated by instanton calculations.

After Witten's work, some examples were explicitly calculated along this
program and further the mathematical justification was established in terms
of operator method.$^{{\rm 4-9}}$

The main concern of this paper is therefore to describe a detailed
construction of the coboundary operator when the manifold is taken to
%p3
%
be the mapping space $\Omega_{p,q}(N)$, which is infinite-dimensional.
In what follows, we are going to
apply Witten's idea to $\Omega_{p,q}(N)$, but before we proceed, let us
clarify the data concerning Morse theory.$^{{\rm 10}}$
\begin{enumerate}
\item[(M-1)] One can define a functional on $\Omega_{p,q}(N)$ as follows,
\beq
E[\phi] = \frac{1}{2}\int_0^1dx\left( \del,\del \right)\ \  {\rm for}\  \phi
\in \Omega_{p,q}(N).
\eeq
This functional will play the role of the Morse function on $\Omega_{p,q}(N)$
when two points $p$ and $q$ on $N$ are not conjugate.
\item[(M-2)] The critical points of $E$ are the geodesics on $N$, where
the gradient of $E$, $\grad = -\nabla_{x}\del$, vanishes.
\item[(M-3)] The Hessian at the geodesic $\phi$ is the self-adjoint
differential operator  $J_{\phi} : \Gamma(\phi^{*}TN)
 \longrightarrow \Gamma(\phi^{*}TN)$ (Jacobi operator) defined by
\beq
J_{\phi} = -\nabla_{x}\nabla_{x}-R\left(\,\bullet\,,\del\right)\del.
\eeq
\end{enumerate}
Here we have used the following conventions;
\ $(\bullet,\bullet)$ is the inner product
with respect to the metric $g$, $\nabla_{x}$ the pull-back of the covariant
derivative on $TN$ and $R$ the Riemannian curvature.$^{{\rm 11}}$

This paper is organized as follows. In the next section we introduce
the action of a topological field theory and discuss the instanton
solutions, which are given by the solutions of a nonlinear heat
equation. In section III, we describe the Hamiltonian formalism; the
cohomology of the mapping space is identified with the BRST cohomology
and the perturbative realization is investigated using
normal coordinates on the mapping space. Section IV is concerned
with the tunneling between the perturbative vacua, in which we calculate
the matrix element of the BRST operator between the vacua, and determine
the coboundary operator of a cohomology complex. In section V we study
a simple example to illustlate our calculations. The final section
is devoted to concluding remarks. Some of the detailed calculations
are presented in the Appendix.

%p4
%
%
%
\section{MODEL}
\setcounter{equation}{0}
According to Langevin approach,$^{{\rm 12}}$ we start with the action:
\beqa
S_{0}&=&\frac{1}{2}\inta
\left(K(\phi,G),\,K(\phi,G)\right),\label{eq:es0} \\
K&=&G-\frac{\partial\phi}{\partial t}+\grad,
\eeqa
where in local coordinates,
\beq
(K,K) = g_{ij}
\left(G^{i}-\frac{\partial\phi}{\partial t}^{i}+\grad^{i}\right)
\left(G^{j}-\frac{\partial\phi}{\partial t}^{j}+\grad^{j}\right),
\eeq
\beq
{\rm with}\ \ \grad^{i} = -\frac{\partial^{2}\phi^{i}}{\partial x^{2}}
-\Gamma^{i}_{jk}\,\del^{j}\del^{k}.
\eeq
Here $\phi$ denotes the mapping $\Sigma \rightarrow N$ and $G\in\Gamma
(\phi^{*}TN)$ the auxiliary field. The above action describes a system with
no dynamical degrees
of freedom: a shift of the field $G$ could eliminate all of the $\phi$
dependences. In such a situation, the Batalin-Vilkovisky algorithm enables
us to construct a BRST invariant quantum action.$^{{\rm 13}}$
A crucial feature of
$S_{0}$ is that it has a local symmetry given by
\beqa
\delta\phi^{i}&=&\epsilon^{i},\nonumber \\
\delta G^{i}&=&\frac{\partial\epsilon^{i}}{\partial t}+
\frac{\partial^{2}\epsilon^{i}}{\partial x^{2}}+
2\Gamma^{i}_{jk}\,\del^{k}\frac{\partial\epsilon^{j}}{\partial x} \\
& & {}+\partial_{j}\Gamma^{i}_{k\ell}\ \del^{k}\del^{\ell}\,\epsilon^{j}-
\Gamma^{i}_{jk}\,K^{k}\epsilon^{j},  \nonumber
\eeqa
where $\epsilon^{i}$ are local infinitesimal parameters of
the transformation.

%p5
%
The algorithm proceeds with standard calculations; here we simply
quote the result. The gauge-fixed Euclidian action is Q-exact:
\beqa
S &=& \inta\,\Bigl\{-\frac{1}{2}\left(B,\,B\right)
-\left(\frac{\partial\phi}{\partial t}-\grad,\,B\right) \nonumber \\
& & {}+\left(\psi,\,\nabla_{t}\psi^{*}-J_{\phi}(\psi^{*})\right)
+\frac{1}{4}\left(\psi,\,R(\psi^{*},\psi^{*})\psi\right)\Bigr\} \nonumber \\
&=& \inta\,Q\left(\psi,\,\frac{\partial\phi}{\partial t}-\grad
+\frac{1}{2}B\right),    \label{eq:action}
\eeqa
with the BRST symmetry,
\beqa
Q\phi^{i}&=&-\psi^{*i},\nonumber \\
Q\psi^{*i}&=&0,\nonumber \\
Q\psi_{i}&=&-B_{i}+\psi_{j}\,\Gamma^{j}_{ik}\,\psi^{*k}, \nonumber \\
QB_{i}&=&-B_{j}\,\Gamma^{j}_{ik}\,\psi^{*k}+\frac{1}{2}
\psi_{k}\,R^{k}{ij\ell}\,\psi^{*j}\psi^{*\ell}.    \label{eq:brs}
\eeqa
Here $B$ and $\psi$ (or $\psi^{*}$) are the auxiliary field and the ghost
field in $\Gamma(\phi^{*}TN)$, respectively, and the covariant derivative,
\beq
\nabla_{t}\psi^{*i} = \frac{\partial\psi^{*i}}{\partial t}+\Gamma^{i}_{jk}\,
\frac{\partial\phi^{j}}{\partial t}\,\psi^{*k},
\eeq
is the pull-back of the covariant derivative on $TN$.

Furthermore, it is easy to see that the action is invariant (up to the
surface term) under the discrete transformation,
\beq
\psi \rightarrow \psi^{*}, \ \psi^{*} \rightarrow \psi,
\ B \rightarrow B-2\grad \label{eq:sym}
\eeq
%p6
%
and
\beq
E[\phi] \rightarrow -E[\phi]\qquad (J_{\phi} \rightarrow -J_{\phi})
\label{eq:sym2}
\eeq
at the same time. As a result, the action Eq.(\ref{eq:action})
permits the second BRST
symmetry $Q^{*}$, which is obtained by substituting
Eqs.(\ref{eq:sym}) and (\ref{eq:sym2})
into Eq.(\ref{eq:brs}),
\beqa
Q^{*}\phi^{i}&=&-\psi^{i}, \nonumber \\
Q^{*}\psi^{i}&=&0,  \nonumber \\
Q^{*}\psi^{*}_{i}&=&-B_{i}+2g_{ij}\,
\grad^{j}+\psi^{*}_{j}\,\Gamma^{j}_{ik}\,
\psi^{k}, \nonumber \\
Q^{*}B_{i}&=&2Q^{*}\left(g_{ij}\,\grad^{j}\right)-B_{j}\,
\Gamma^{j}_{ik}\,
\psi^{k} \nonumber \\
& & {}+2g_{jk}\,\grad^{k}\,\Gamma^{j}_{i\ell}\,\psi^{\ell}+\frac{1}{2}
\psi^{*}_{k}\,R^{k}_{ij\ell}\,\psi^{j}\,\psi^{\ell}.
\eeqa
Using the equation of motion \,$\delta S/\delta B_{i} = 0$, i.e.,
\beq
B^{i} = - \frac{\partial\phi^{i}}{\partial t}+\grad^{i},
\eeq
we eliminate $B$ and the action takes the form,
\beqa
S&=&\inta\,\Bigl\{\frac{1}{2}\left(\frac{\partial\phi}{\partial t},\,
\frac{\partial\phi}{\partial t}\right)+
\frac{1}{2}\left(\grad,\,\grad\right)
\nonumber \\
& & {}+ \left(\psi,\,\nabla_{t}\,\psi^{*}-J_{\phi}(\psi^{*})\right)+
\frac{1}{4}\left(\psi,\,R(\psi^{*},\psi^{*})\psi\right)\Bigr\}.
 \label{eq:act2}
\eeqa

Before going into the quantization, we here consider
 the classical solutions
(instantons) corresponding to tunneling processes.
 The geodesics on $N$,
%p7
%
where $\grad = 0$, can be regarded as the degenerate
minima of the potential
term,
\beq
V[\phi] = \frac{1}{2}\int_0^1dx\,\left(\grad,\,\grad\right).
\eeq
Therefore, the relevant instantons are the steepest descent
 paths leading from
one geodesic to another geodesic. They would be extrema of
 the bosonic part of
the action,
\beqa
S_{B}&=&\inta\,\Bigl\{\frac{1}{2}\left(
\frac{\partial\phi}{\partial t},\,\frac{\partial\phi}{\partial t}\right)
+\frac{1}{2}\left(\grad,\,\grad\right)\Bigr\}
\nonumber \\
&=&\inta\,\Bigl\{\frac{1}{2}\left(\frac{\partial\phi}{\partial t}\mp\grad,\,
\frac{\partial\phi}{\partial t}\mp\grad\right)
\nonumber \\
& & {}\pm\left(\frac{\partial\phi}{\partial t},\,\grad\right)\Bigr\}.
\eeqa

Let us fix the surface term in the above equation: $\phi(t,x)\in
\Omega_{p,q}(N)$ for any fixed $t$.
Then, the second term in $S_{B}$ is rewritten
as follows:
\beqa
\inta\,\left(\delt,\,\grad\right)
&=&\inta\,\Bigl\{-\frac{d}{dx}\left(\delt,\del\right)+\frac{1}{2}
\frac{d}{dt}\left(\del,\del\right)\Bigr\} \nonumber \\
&=&E[\phi(t=\infty)]-E[\phi(t=-\infty)],
\eeqa
and hence we obtain an inequality,
\beq
S_{B} \ge |\ E[\phi(t=\infty)]-E[\phi(t=-\infty)]\ |.
\eeq
%p8
%
The configurations that minimize the action are given by the solutions of
a nonlinear heat equation,
\beq
\delt = \pm\,\grad,   \label{eq:grad}
\eeq
or in terms of the local coordinates,
\beq
\delt^{i} = \mp\left(\frac{\partial^{2}\phi^{i}}{\partial x^{2}}
+\Gamma^{i}_{jk}\frac{\partial\phi^{j}}{\partial x}
\frac{\partial\phi^{k}}{\partial x}\right).     \label{eq:ins}
\eeq
In particular, the solutions connecting geodesics express
the steepest descent
paths, and we call positive (negative) gradient flows instantons
(anti-instantons).

The next step is to investigate the space of these solutions.
However, the direct study of the nonlinear heat equation is quite difficult
and in general one can only prove that the instantons exist
globally.$^{{\rm 14,15}}$
Fortunately, for our purpose, the explicit form of the solutions
is not necessary.

For each geodesic $\gamma$ with the Morse index $\ell_{\gamma}$
in $\Omega_{p,q}(N)$, we associate a variety
$V^{(+)}_{\gamma}$ $(V^{(-)}_{\gamma})$, which is a union of the
positive gradient solutions of Eq.(\ref{eq:grad})
departing from (arriving at) $\gamma$; $V^{(+)}_{\gamma}$
($V^{(-)}_{\gamma}$) is a subvariety of codimension $\ell_{\gamma}$
(dimension $\ell_{\gamma}$) contained in $\Omega_{p,q}(N)$.
To see local expression for the varieties, let us consider the tangent
space at $\gamma$. Using the exponential mapping,
$\ {\rm exp}_{\gamma(x)}\,: T_{\gamma(x)}N \longrightarrow N$,
we write the general solution of Eq.(\ref{eq:ins}) near $\gamma$
as follows,
\beq
\phi(t,x) = {\rm exp}_{\gamma(x)}\xi(t,x) ,\ \ \xi(t,x)
\in T_{\gamma(x)}N.
\label{eq:mayurin}
\eeq
Substituting Eq.(\ref{eq:mayurin}) into Eq.(\ref{eq:ins}) and looking to the
first order in $\xi$, we see that $\xi$ must satisfy
%p9
%
\beq
\frac{\partial\xi}{\partial t}(t,x) = J_{\gamma}\xi(t,x),
\label{eq:eign}
\eeq
with the condition $\xi(t,0)=\xi(t,1)=0$ for all $t$.
If we decompose $\xi$ into eigenfunctions of $J_{\gamma}$,
\beq
\xi(t,x) = \sum_{\alpha=0}^{\infty} \xi^{\alpha}(t)e_{\alpha}(x),\ \
J_{\gamma}e_{\alpha}(x) = \lambda_{\alpha}e_{\alpha}(x),
\eeq
Eq.(\ref{eq:eign}) reduces to
\beq
\frac{d\xi^{\alpha}}{dt}(t) = \lambda_{\alpha}\,\xi^{\alpha}(t)
\qquad\mbox{for all}\ \alpha.
\eeq
These equations can be easily solved and determine the varieties
$V^{(\pm)}_{\gamma}$ near $\gamma$:
\beqa
V^{(+)}_{\gamma}&=&{\rm exp}_{\gamma(x)}
\Bigl(\sum_{\lambda_{\alpha}>0}\,c^{(+)}_{\alpha}\,
e^{\lambda_{\alpha}\,t}\,e_{\alpha}(x)\Bigr)
\ \ {\rm for}\ t\,\rightarrow\,-\infty,\\
V^{(-)}_{\gamma}&=&{\rm exp}_{\gamma(x)}
\Bigl(\sum_{\lambda_{\alpha}<0}\,c^{(-)}_{\alpha}\,
e^{\lambda_{\alpha}\,t}\,e_{\alpha}(x)\Bigr)
\ \ {\rm for}\ t\,\rightarrow\,\infty,
\eeqa
with constants $c^{(\pm)}_{\alpha}$.

Let $\gamma_{A}$ and $\gamma_{B}$ be two geodesics with the Morse indices
$\ell$ and $\ell+1$, respectively. Then, our assumptions are following:
\begin{enumerate}
\item[(A-1)] There exist finite number of instantons connecting $\gamma_{A}$
with $\gamma_{B}$.
\item[(A-2)] $V^{(+)}_{\gamma_{A}}$ and $V^{(-)}_{\gamma_{B}}$ transversely
intersect along these instantons.
\end{enumerate}
%p1o
%
Under these assumptions we write  the moduli space of the instantons as,
\beq
{\cal M}_{AB} = \bigcup_{intersection}\ \left(V^{(+)}_{\gamma_{A}}\,
\cap\,V^{(-)}_{\gamma_{B}} \right)
\eeq
and deduce the dimension 1.
\begin{enumerate}
\item[(A-3)] The tangent spaces along each instanton satisfy the
transversality condition. This will be specified in section IV.
\end{enumerate}
\section{HAMILTONIAN FORMALISM}
\setcounter{equation}{0}
In this section we discuss the perturbative vacua for the system. This will
be done in the Hamiltonian framework.

The canonical momenta and the Poisson
bracket are readily read off from Eq.(\ref{eq:act2}),
\beq
p_{i}=\frac{\delta^{r}S}{\delta\left(\delt^{i}\right)}
=g_{ij}\delt^{j}-\Gamma^{k}_{ij}\psi^{*j}\psi_{k}, \ \ \
\psi_{i}=\frac{\delta^{r}S}{\delta
\left(\frac{\partial\psi^{*i}}{\partial t}\right)}
\eeq
and
\beq
\bigl\{\phi^{i}(x),\,p_{j}(x^{\prime})\bigr\}
=\delta^{i}_{j}\,\delta(x-x^{\prime}),\ \ \
\bigl\{\psi^{*i}(x),\,\psi_{j}(x^{\prime})\bigr\}
=\delta^{i}_{j}\,\delta(x-x^{\prime}). \label{eq:poi}
\eeq
Then, the Hamiltonian corresponding to $S$ can be written as follows,
\beqa
H&=&\int_0^1dx\,\Bigl\{\frac{1}{2}\left(\pi,\,\pi\right)-\frac{1}{2}
\left(\grad,\,\grad\right)  \nonumber \\
& & {} +\left(\psi,\,J_{\phi}(\psi^{*})\right)
-\frac{1}{4}\left(\psi,\,R(\psi^{*},\psi^{*})\psi\right)\Bigr\},
\label{eq:hami}
\eeqa
%p11
%
where $\pi_{i} = p_{i}+\Gamma^{k}_{ij}\,\psi^{*j}\,\psi_{k}$.
The BRST charges are obtained by the Noether prescription and their action
is given on using the Poisson bracket Eq.(\ref{eq:poi}). In fact we find
\beqa
Q&=&\int_0^1dx\,\left(\psi^{*},\,\pi-\grad\right),  \label{eq:q} \\
Q^{*}&=&\int_0^1dx\,\left(\psi,\,\pi+\grad\right), \label{eq:qq}
\eeqa
which satisfy the nilpotent condition,
\beq
\bigl\{Q,\,Q\bigr\}=0,\ \ \bigl\{Q^{*},\,Q^{*}\bigr\}=0.
\eeq
The Hamiltonian Eq.(\ref{eq:hami}) may be rewritten as
\beq
H=\frac{1}{2}\bigl\{Q,\,Q^{*}\bigr\},
\eeq
which is familiar with supersymmetric quantum theories.

Now let us proceed to the quantization. The classical phase space will be
identified with the cotangent bundle of
 $\displaystyle \bigoplus_{r}\,
 \Lambda^{r}\,\left(T^{*}\Omega_{p,q}(N)\right)$.
So the quantum Hilbert space ${\cal H}$ is the space
 of form-valued functions,
i.e., $\displaystyle {\cal H} =
\Gamma\,\Bigl(\bigoplus_{r}\,
\Lambda^{r}\,\left(T^{*}\Omega_{p,q}(N)\right)\Bigr)$.
Then the coordinate
operators $\widehat\phi^{i}$ and $\widehat\psi^{*i}$ act as multiplication
and exterior multiplication on the wave functions, respectively, while
the momentum operators $\widehat p_{i}$ and $\widehat\psi_{i}$ act by
differentiation, $\widehat p_{i} \sim \delta/\delta\phi^{i}$ and
$\widehat\psi_{i} \sim \delta/\delta\psi^{*i}$ (interior product
on forms).

%p12
%
We consider a local representation of the quantum Hilbert space.
Let us first introduce a local coordinate system in $\Omega_{p,q}(N)$. For
$\phi_{0}\in \Omega_{p,q}(N)$ an exponential mapping,
\beq
{\rm Exp}_{\phi_{0}}\ :\ T_{\phi_{0}}\,\Omega_{p,q}(N)=
\Gamma\left(\phi^{*}_{0}\,TN\right)\ \longrightarrow\ \Omega_{p,q}(N),
\eeq
is defined by ${\rm Exp}_{\phi_{0}}\,(X) =
{\rm exp}\circ X$, where exp denotes
the usual exponential mapping on $N$ and the symbol $\lq\lq\circ$" the
composition of mappings, i.e., for
\ $X\in\Gamma\left(\phi^{*}_{0}\,TN\right)$,
\ $X(x)\in T_{\phi_{0}(x)}\,N$ \,so that
$\exp\circ X(x) = \exp_{\phi_{0}(x)}\,X(x)\in N$.
Using an orthonormal basis $e_{\alpha}\ (\alpha=0,1,...)$ of
$\Gamma\left(\phi^{*}_{0}\,TN\right)$,
\beqa
\langle e_{\alpha},\,e_{\beta} \rangle&=&\int_0^1dx\,\left(e_{\alpha}(x),\,
e_{\beta}(x)\right) \nonumber \\
&=& \delta_{\alpha\beta},
\eeqa
we have a normal coordinate system $\phi = (\xi^{0},\,\xi^{1},...)$
with origin $\phi_{0}$ in $\Omega_{p,q}(N)$,
\beq
\phi={\rm Exp}_{\phi_{0}}\,(\xi)
=\exp\circ\sum_{\alpha=0}^\infty\,\xi^{\alpha}\,e_{\alpha},\ \
\xi\in\Gamma\left(\phi^{*}_{0}\,TN\right).
\eeq
Thus any wave function in ${\cal H}$ may be locally written in the form,
\beqa
\omega&=&\bigoplus_{r}\,\omega^{(r)},\\
\omega^{(r)}&=&\sum_{\alpha_{1}<\cdots<\alpha_{r}}\,
\omega_{\alpha_{1}\cdots\alpha_{r}}\,(\xi)\,\psi^{*\alpha_{1}}\cdots
\psi^{*\alpha_{r}}\ \in \Gamma\left(\Lambda^{r}\,T^{*}\Omega_{p,q}(N)\right),
\eeqa
where we have identified $d\xi^{\alpha}$ with $\psi^{*\alpha}$.

%p13
%
The inner product on $\Gamma\left(\phi^{*}\,TN\right)$, in the normal
coordinate system $\phi = {\rm Exp}_{\phi_{0}}\left(\xi\right)$,
is expressed as
\beqa
G_{\alpha\beta}&=&\Bigl\langle \frac{\partial}{\partial\xi^{\alpha}},\,
\frac{\partial}{\partial\xi^{\beta}} \Bigr\rangle  \\
&=&\int_0^1dx\,
\left(\frac{\partial}{\partial\lambda}\left(\exp\circ(\xi
 + \lambda e_{\alpha})
(x)\right)\Big|_{\lambda=0},\,\frac{\partial}{\partial\lambda}\left(
\exp\circ(\xi + \lambda e_{\beta})(x)\right)
\Big|_{\lambda=0}\right), \nonumber
\eeqa
and the expansion of $G_{\alpha \beta}$ takes the form (Appendix),
\beq
G_{\alpha \beta} = \delta_{\alpha \beta}-\frac{1}{3}\int_0^1dx\,
\left(R(e_{\alpha},\,\xi)\xi,\,e_{\beta}\right)+{\cal O}(\xi^{3}).
\label{eq:proof1}
\eeq
Then $G^{\alpha \beta} = G^{-1}_{\alpha \beta} = \langle \psi^{*\alpha}
,\,\psi^{*\beta}\rangle$, which may be extended to
${\cal H}$ : First define a product,
\beq
\langle \psi^{*\alpha_{1}}\cdots\psi^{*\alpha_{p}},\,\psi^{*\beta_{1}}\cdots
\psi^{*\beta_{p}} \rangle = \sum_{r_{1},...,r_{p}}
\epsilon_{r_{1}\cdots r_{p}}\,G^{\alpha_{1}\beta_{r_{1}}}\cdots
G^{\alpha_{p}\beta_{r_{p}}},
\label{eq:product}
\eeq
where $(r_{1},...,r_{p})$ runs over $(1,...,p)$ and
\beq
\epsilon_{r_{1}\cdots r_{p}}=1\,(-1)\quad {\rm if}\,
\left({1\,,\cdots,\,p \atop r_{1},\cdots,r_{p}}\right) {\rm is\ even
\,(odd)}.
\eeq
It follows that
\beq
\langle \omega^{(r)},\,\eta^{(r)} \rangle =
\sum_{\alpha_{1}<\cdots<\alpha_{r}}
\,\omega_{\alpha_{1}\cdots\alpha_{r}}\,(\xi)\,
\eta^{\alpha_{1}\cdots\alpha_{r}}\,(\xi)
\eeq
\beq
\qquad {\rm for}\ \omega^{(r)},\ \eta^{(r)}\,\in
\Gamma\left(\Lambda^{r}\,T^{*}\Omega_{p,q}(N)\right), \nonumber
\eeq
and a scalar product on ${\cal H}$ is formally given by
\beq
\langle \omega^{(r)}|\,\eta^{(s)} \rangle = \delta_{rs}\,
\int\prod_{\alpha}d\xi^{\alpha}\,\sqrt{G}\,\langle \omega^{(r)},
\,\eta^{(s)} \rangle.
\eeq

%p14
%
Now we discuss the BRST operators.
These operators, which are quantum versions
of Eqs.(\ref{eq:q}) and (\ref{eq:qq}), will be written in the form,
\beqa
\widehat Q &=& \sum_{\alpha=0}^{\infty}\,\widehat\psi^{*\alpha}\,
\left(\nabla_{\frac{\partial}{\partial\xi^{\alpha}}}+\grad_{\alpha}
\right), \\
\widehat Q^{*} &=& \sum_{\alpha=0}^{\infty}\,G^{\alpha \beta}(\xi)
\widehat\psi_{\beta}\,
\left(-\nabla_{\frac{\partial}{\partial\xi^{\alpha}}}+\grad_{\alpha}
\right).
\eeqa
Here $\nabla_{\frac{\partial}{\partial\xi^{\alpha}}}$ is the covariant
derivative associated with the metric $G$ and
\beqa
\grad_{\alpha}&=&\Bigl\langle \grad,\,\frac{\partial}{\partial\xi^{\alpha}}
\Bigr\rangle  \nonumber \\
&=&\int_0^1dx\,\left(-\nabla_{x}\,\del,\,\frac{\partial}{\partial\lambda}
\left(\exp\circ(\xi+\lambda\,e_{\alpha})(x)\right)\Big|_{\lambda=0}\right).
\label{eq:gg}
\eeqa
Then, we have the following expansion (Appendix):
\beq
\grad_{\alpha}=\int_0^1dx\,\bigl\{\left(\sgrad\,E[\phi_{0}]
,\,e_{\alpha}\right)+\left(J_{\phi_{0}}(\xi),\,e_{\alpha}\right) \bigr\}
+{\cal O}(\xi^{2}).  \label{eq:proof2}
\eeq
By introducing exterior differential $d$ and formal adjoint $d^{*}$,
\beqa
d&=&\sum_{\alpha=0}^{\infty}\widehat\psi^{*\alpha}\nabla_{\frac{\partial}
{\partial \xi^{\alpha}}}
\ :\ \Gamma(\Lambda^{r}T^{*}\Omega_{p,q}(N))\ \rightarrow\
\Gamma(\Lambda^{r+1}T^{*}\Omega_{p,q}(N)),  \\
d^{*}&=&-\sum_{\alpha=0}^{\infty}\widehat\psi^{\alpha}
\nabla_{\frac{\partial}{\partial \xi^{\alpha}}}
\ :\ \Gamma(\Lambda^{r+1}T^{*}\Omega_{p,q}(N))\ \rightarrow\
\Gamma(\Lambda^{r}T^{*}\Omega_{p,q}(N)),
\eeqa
the BRST operators is conveniently expressed as follows,
\beqa
\widehat Q &=& e^{-E[\phi]}\,d\,e^{E[\phi]},  \\
\widehat Q^{*}&=& e^{E[\phi]}\,d^{*}\,e^{-E[\phi]},
\eeqa
%p15
%
and hence the quantum Hamiltonian is given by
\beq
2\widehat H = \widehat Q\widehat Q^{*}+\widehat Q^{*}\widehat Q.
\eeq
These operators may be considered as an infinite-dimensional extension
of Witten's cohomology operators in supersymmetric quantum mechanics on
finite-dimensional manifolds.$^{{\rm 3}}$

For the remainder of this section, we will discuss the physical Hilbert
space ${\cal H}_{phys}$. This is usually defined by the BRST
cohomology: the condition $\widehat Q\,\omega_{phys}
\,=\,0$,\  $\omega_{phys}
\,\in\,{\cal H}$ and furthermore $\omega^{'}_{phys}\,=\,\omega_{phys}\,+
\widehat Q\,(\,\bullet\,)$ is equivalent to $\omega_{phys}$.
By using the following complex,
\begin{eqnarray*}
0\quad\longrightarrow\quad\Lambda^{0}\,T^{*}\Omega_{p,q}(N)\quad
\stackrel{d}{\longrightarrow}\quad\Lambda^{1}\,T^{*}\Omega_{p,q}(N)\quad
\stackrel{d}{\longrightarrow}\quad\cdots
\end{eqnarray*}
\beq
\Big\downarrow\quad e^{-E[\phi]}\qquad\qquad\qquad
\Big\downarrow\quad e^{-E[\phi]}
\eeq
\begin{eqnarray*}
0\quad\longrightarrow\quad\Lambda^{0}\,T^{*}\Omega_{p,q}(N)\quad
\stackrel{\widehat Q}{\longrightarrow}\quad
\Lambda^{1}\,T^{*}\Omega_{p,q}(N)\quad
\stackrel{\widehat Q}{\longrightarrow}\quad\cdots ,
\end{eqnarray*}
${\cal H}_{phys}$ is expressed as follows:
\beqa
{\cal H}_{phys}&=&{\rm Ker}\,\widehat Q/{\rm Im}\,\widehat Q  \nonumber  \\
&\simeq&{\rm Ker}\,\widehat H    \nonumber  \\
&\simeq&{\rm Ker}\,d/{\rm Im}\,d.
\eeqa
%p16
%
These isomorphisms  show that ${\cal H}_{phys}$ is the vector space of
quantum vacua  or equivalently
the de Rham cohomology $H^{*}(\Omega_{p,q}(N))$.

We now investigate the quantum vacua by a perturbative approximation.
This is done by taking a normal coordinate system around
each geodesic, which
corresponds to the classical vacuum.
 Let $\gamma$ be a geodesic with the Morse
index $\ell$ and $e_{\alpha}\in\Gamma(\gamma^{*}TN)\,(\alpha=0,1,...)$
a basis consisting of the orthonormal eigenfunctions of $J_{\gamma}$
with the eigenvalues
$\lambda_{\alpha}<0$ ($\alpha=0,1,...,\ell-1$) and $\lambda_{\alpha}
>0$ ($\alpha\ge\ell$).
Using the expansion formulas Eqs.(\ref{eq:proof1})
 and (\ref{eq:proof2}),
applied to the normal coordinate system $\phi={\rm Exp}_{\gamma}
\left(\sum_{\alpha=0}^{\infty}\xi^{\alpha}e_{\alpha}\right)$ where
${\rm grad}\,E[\gamma]=0$ and $J_{\gamma}(e_{\alpha})=\lambda_{\alpha}
e_{\alpha}$, we have approximate BRST operators in a neighborhood
of $\gamma$,
\beqa
\widehat Q_{\gamma}&=&\sum_{\alpha=0}^{\infty}\,\widehat\psi^{*\alpha}
\left(\frac{\partial}{\partial\xi^{\alpha}}+\lambda_{\alpha}\xi^{\alpha}
\right), \\
\widehat Q_{\gamma}^{*}&=&\sum_{\alpha=0}^{\infty}\,\widehat\psi_{\alpha}
\left(-\frac{\partial}{\partial\xi^{\alpha}}+\lambda_{\alpha}\xi^{\alpha}
\right),
\eeqa
It readily follows that the corresponding Hamiltonian is
\beq
\widehat H_{\gamma}=\sum_{\alpha=0}^{\infty}\,
\left(-\left(\frac{\partial}{\partial\xi^{\alpha}}\right)^{2}+
\left(\lambda_{\alpha}\xi^{\alpha}\right)^{2}+\lambda_{\alpha}
[\widehat\psi^{*\alpha},\,\widehat\psi_{\alpha}]\right)
\eeq
and the normalized vacuum state of $\widehat H_{\gamma}$, zero-eigenvalue
function, is given by
\beq
\Omega_{\gamma}=\prod_{\alpha}\,
\left(\frac{|\lambda_{\alpha}|}{\pi}\right)^
{\frac{1}{4}}\,e^{-\frac{1}{2}|\lambda_{\alpha}|(\xi^{\alpha})^{2}}\,
\omega^{(\ell)}[\varepsilon],
\eeq
%p17
%
where $\omega^{(\ell)}[\varepsilon]$ is the $\ell$-form with
negative-eigenvalue indices in the frame $\varepsilon=(e_{0},e_{1},...)$,
\beq
\omega^{(\ell)}[\varepsilon]=\psi^{*0}\psi^{*1}\cdots\psi^{*\ell-1}.
\label{eq:orient}
\eeq

Let $C^{\ell}=\{\gamma_{I}\,|\,I=1,2,...,n(\ell)\}$ be a set of geodesics
with the Morse index $\ell$. For each $\gamma_{I}\in C^{\ell}$,
we can assign a $\ell$-form vacuum $\Omega_{\gamma_{I}}$,
which induces an orientation
$\omega^{(\ell)}[\varepsilon^{I}]$ on the $\ell$-dimensional vector
space $N(\gamma_{I})\subset\Gamma\left(\gamma_{I}^{*}TN\right)$
spanned by the negative-eigenvalue functions of $J_{\gamma_{I}}$.
Then, ${\cal H}_{phys}$ is approximated by
\beq
{\cal H}_{phys}\simeq\bigoplus_{\ell=0}^{\infty}\,X^{\ell},
\label{eq:hphys}
\eeq
where $X^{\ell}$ is the $n(\ell)$-dimensional vector space
spanned by $\Omega_{\gamma_{I}}$.
\section{TUNNELING CALCULATIONS}
\setcounter{equation} {0}
In this section, we evaluate transition matrix elements between
the perturbative vacua. These can be used to construct the coboundary
operator $\delta^{(\ell)}\,:\,X^{\ell}\,\longrightarrow\,X^{\ell+1}$.

Suppose we have an instanton $\phi_{0}$
connecting two geodesics $\gamma_{A}$
and $\gamma_{B}$ with the Morse indices $\ell$ and $\ell+1$,
 respectively.
As we saw in section II, the instanton has exactly one free parameter
corresponding to the time translation. Then the vector field
%
%p18
%
$\frac{\partial\phi_{0}}{\partial t}$ is in the kernel of the operator,
\beq
\nabla_{t}-J_{\phi_{0}}\,:\,\Gamma(\phi^{*}_{0}\,TN)\,\longrightarrow\,
\Gamma(\phi^{*}_{0}\,TN)
\eeq
and
\beq
\inta\,\left(\frac{\partial\phi_{0}}{\partial t},\,
\frac{\partial\phi_{0}}{\partial t}\right)=
E[\gamma_{B}]-E[\gamma_{A}]<\infty.
\eeq
This means that there exists one ghost zero mode, which forces the
partition function to vanish. An analogous situation appeared in
supersymmetric quantum mechanics.$^{{\rm 3-8}}$
So if we wish to calculate
the nontrivial
transition between the perturbative vacua, we must take the matrix element
$\langle \Omega_{\gamma_{B}}|\,\widehat Q\,|\Omega_{\gamma_{A}}\rangle$,
in which $\widehat Q$ absorbs the zero mode in the ghost path
integral measure.

Using the commutation relation $[\widehat Q,\,f[\phi]\,]=df[\phi]$ for any
functional $f[\phi]$, and the fact $\langle \Omega_{\gamma_{I}}|\,df[\phi]
\,|\Omega_{\gamma_{I}}\rangle\simeq f[\gamma_{I}]\quad(I=A,B)$, we get
\beq
\langle \Omega_{\gamma_{B}}|\,\widehat Q\,|\Omega_{\gamma_{A}}\rangle
\simeq \frac{1}{f[\gamma_{A}]-f[\gamma_{B}]}\,\langle \Omega_{\gamma_{B}}|
\,df[\phi]\,|\Omega_{\gamma_{A}}\rangle.
\label{eq:mat}
\eeq
We now use standard path integral techniques. Let us consider the
heat kernel with the insertion $df$,
\beq
K_{df}[\gamma_{A}\,\psi_{A},\,\gamma_{B}\,\psi^{*}_{B}]\in {\rm Hom}\left(
\Lambda^{\ell}\,T^{*}_{\gamma_{A}}\Omega_{p,q}(N),\,
\Lambda^{\ell+1}\,T^{*}_{\gamma_{B}}\Omega_{p,q}(N)\right),
\eeq
defined by the path integral
\beq
K_{df}[\gamma_{A}\,\psi_{A},\,\gamma_{B}\,\psi^{*}_{B}]=
\int\,[d\phi][d\psi^{*}][d\psi]\,df[\phi]\,e^{-S},
\eeq
%
%p19
%
where the boundary conditions are
\beqa
\phi(-\infty,\,x) &=& \gamma_{A}(x),\quad \phi(\infty,\,x)=\gamma_{B}(x), \\
\psi(-\infty,\,x) &=& \psi_{A}(x)\quad {\rm and}\quad \psi^{*}(\infty,\,x)
=\psi^{*}_{B}(x). \label{eq:bc}
\eeqa

The quantum effect of interest is the quadratic fluctuation around each
instanton. We denote by $\phi_{s}$ a small deformation from an instanton
$\phi$,
\beq
\phi_{s}(t,x)={\rm exp}_{\phi(t,x)}\,s\xi(t,x),
\eeq
where $s$ is the deformation parameter and
\beq
\phi_{0}(t,x)=\phi(t,x),\quad \frac{\partial\phi_{s}}{\partial s}(t,x)\Big|
_{s=0}=\xi(t,x)\,\in\,T_{\phi(t,x)}N.
\eeq
Using the expansion,
\beqa
\left(\frac{\partial\phi_{s}}{\partial t}-{\rm grad}\,E[\phi_{s}],
\,\frac{\partial\phi_{s}}{\partial t}-{\rm grad}\,E[\phi_{s}]\right)
\nonumber  \\
=s^{2}\left(\nabla_{t}\xi-J_{\phi}(\xi),\,\nabla_{t}\xi-J_{\phi}(\xi)
\right)+{\cal O}(s^{3}),
\eeqa
we obtain the action up to the quadratic order\,:
\beqa
S&=&E[\gamma_{B}]-E[\gamma_{A}]+S^{(2)}+\,\mbox{higher order}, \\
S^{(2)}&=&\int_{-T}^{T}dt\,\int_{0}^{1}dx\,\Bigl\{\frac{1}{2}
\left(\nabla_{t}\xi-J_{\phi}(\xi),\,\nabla_{t}\xi-J_{\phi}(\xi)
\right) \nonumber  \\
& & {}+\left(\psi,\,\nabla_{t}\psi^{*}-J_{\phi}(\psi^{*}\right)\Bigr\}
\quad(T\,\rightarrow\,\infty).  \label{eq:aact2}
\eeqa

The treatment of the path integral may be further simplified if
%
%p20
%
the adiabatic approximation applies to Eq.(\ref{eq:aact2}). For this
we change the integral variable $t=T\tau$ and $\xi(t,x)=\sqrt{T}
\xi(\tau,x)$ at the same time, so that $S^{(2)}$ becomes
\beqa
S^{(2)}&=&\int_{-1}^{1}d\tau\,\int_{0}^{1}dx\,\Bigl\{\frac{1}{2}
\left(\nabla_{\tau}\xi-TJ_{\phi}(\xi),\,\nabla_{\tau}\xi-TJ_{\phi}(\xi)
\right) \nonumber  \\
& & {}+\left(\psi,\,\nabla_{\tau}\psi^{*}-TJ_{\phi}(\psi^{*}\right)\Bigr\}.
\label{eq:act22}
\eeqa
Let us expand the tangent vectors $\xi$, $\psi$ and $\psi^{*}$ in
terms of the eigenfunctions of $J_{\phi}$,
\beqa
\xi(\tau,x)&=&\sum_{\alpha=0}^{\infty}\,
\xi^{\alpha}(\tau)\,u^{T}_{\alpha}(\tau,x),
\label{eq:eigen1}  \\
\psi(\tau,x)&=&\sum_{\alpha=0}^{\infty}\,
\psi^{\alpha}(\tau)\,u^{T}_{\alpha}(\tau,x),
\label{eq:eigen2} \\
\psi^{*}(\tau,x)&=&\sum_{\alpha=0}^{\infty}\,
\psi^{*\alpha}(\tau)\,u^{T}_{\alpha}(\tau,x), \label{eq:eigen3}
\eeqa
where
\beq
J_{\phi}\,u^{T}_{\alpha}(\tau,x)=\lambda^{T}_{\alpha}
(\tau)\,u^{T}_{\alpha},
\quad u^{T}_{\alpha}(\tau,x)\in T_{\phi(T\tau,x)}N
\eeq
with the normalization
\beq
\int_{0}^{1}dx\,\left(u^{T}_{\alpha},\,u^{T}_{\beta}\right)
=\delta_{\alpha \beta}.
\eeq
A perturbative analysis yields that the eigenfunctions
$u^{T}_{\alpha}$ and the corresponding eigenvalues
$\lambda^{T}_{\alpha}$ have the behavior in the large limit $T$,
%
%p21
%
\beq
u^{T}_{\alpha}(\tau,x)\sim \left\{
\begin{array}{@{\,}ll}
u^{A}_{\alpha}(x)+\displaystyle \sum_{\beta \neq \alpha}
\biggl(\frac{1}{\lambda^{A}_{\beta}-\lambda^{A}_{\alpha}}
\displaystyle \sum_{\lambda^{A}_{\gamma}>0}
\,c^{A}_{\alpha \beta \gamma}\,
e^{\lambda^{A}_{\gamma}T\tau} \biggr)u^{A}_{\beta}(x)
& {\rm for}\quad\tau<0,  \\
u^{B}_{\alpha}(x)+\displaystyle \sum_{\beta \neq \alpha}
\biggl(\frac{1}{\lambda^{B}_{\beta}-\lambda^{B}_{\alpha}}
\displaystyle \sum_{\lambda^{B}_{\gamma}<0}
\,c^{B}_{\alpha \beta \gamma}\,
e^{\lambda^{B}_{\gamma}T\tau}\biggr)u^{B}_{\beta}(x)
& {\rm for}\quad\tau>0,
\end{array}
\right.
\eeq
\beq
\lambda^{T}_{\alpha}(\tau)\sim \left\{
\begin{array}{@{\,}ll}
\lambda^{A}_{\alpha}+\displaystyle
\sum_{\lambda^{A}_{\gamma}>0}
\,c^{A}_{\alpha \alpha \gamma}\,
e^{\lambda^{A}_{\gamma}T\tau}
&\qquad {\rm for}\quad\tau<0, \\
\lambda^{B}_{\alpha}+\displaystyle
\sum_{\lambda^{B}_{\gamma}<0}
\,c^{B}_{\alpha \alpha \gamma}\,
e^{\lambda^{B}_{\gamma}T\tau}
&\qquad {\rm for}\quad\tau>0,
\end{array}
\right.   \label{eq:lam}
\eeq
where $u^{I}_{\alpha}\,(\alpha=0,1,...\,;\,I=A,B)$ are
the eigenfunctions of
$J_{\gamma_{I}}$ with the eigenvalues $\lambda^{I}_{\alpha}$ and
\beqa
c^{I}_{\alpha \beta \gamma}&=&\int_{0}^{1}dx\,
\biggl\{
\biggl(
\Bigl(\nabla_{u^{I}_{\gamma}}R \Bigr)\Bigl(u^{I}_{\alpha},\,
\frac{d\gamma_{I}}{dx}\Bigr)\frac{d\gamma_{I}}{dx},\,u^{I}_{\beta}
\biggr)
+
\biggl(
\Bigl(\nabla_{x}R \Bigr)\Bigl(u^{I}_{\gamma},\,
\frac{d\gamma_{I}}{dx}\Bigr)u^{I}_{\alpha},\,u^{I}_{\beta}
\biggr)
\nonumber \\
& & {} +
2\biggl(
R\Bigl(u^{I}_{\alpha},\,
\frac{d\gamma_{I}}{dx}\Bigr)\nabla_{x}u^{I}_{\gamma},\,u^{I}_{\beta}
\biggr)
+
2\biggl(
R\Bigl(u^{I}_{\gamma},\,
\frac{d\gamma_{I}}{dx}\Bigr)\nabla_{x}u^{I}_{\alpha},\,u^{I}_{\beta}
\biggr)
\biggr\},
\eeqa
which are T-independent constants.

By substituting Eqs.(\ref{eq:eigen1}), (\ref{eq:eigen2})
 and (\ref{eq:eigen3})
into Eq.(\ref{eq:act22}), the quadratic
action takes the form
\beqa
S^{(2)}&=&\int_{-1}^{1}d\tau\,
\biggl\{\frac{1}{2}\sum_{\alpha}
\biggl(
\frac{d\xi^{\alpha}}{d\tau}+\sum_{\beta}\xi^{\beta}
\omega^{T}_{\beta \alpha}
-T\lambda^{T}_{\alpha}\,\xi^{\alpha}
\biggr)^{2}  \nonumber  \\
& & {}+\sum_{\alpha}\psi^{\alpha}
\biggl(\frac{d\psi^{*\alpha}}{d\tau}
+\sum_{\beta}\psi^{*\beta}\omega^{T}_{\beta \alpha}-T\lambda^{T}_{\alpha}
\,\psi^{*\alpha}
\biggr)
\biggr\}.
\label{eq:act222}
\eeqa
Here the connection $\omega^{T}_{\alpha \beta}$ is defined by
\beq
\omega^{T}_{\alpha \beta}(\tau)=\int_{0}^{1}dx\,\left(
\nabla_{\tau}u^{T}_{\alpha}(\tau,x),\,u^{T}_{\beta}(\tau,x)\right)=
-\omega^{T}_{\beta \alpha}(\tau),
\eeq
which behaves for large $T$ like
\beq
\omega^{T}_{\alpha \beta}(\tau)\sim \left\{
\begin{array}{@{\,}ll}
\displaystyle \frac{1}{\lambda^{A}_{\beta}-\lambda^{A}_{\alpha}}
\displaystyle \sum_{\lambda^{A}_{\gamma}>0}
\,c^{A}_{\alpha \beta \gamma}\,\lambda^{A}_{\gamma}T
\,e^{\lambda^{A}_{\gamma}T\tau}
&\qquad {\rm for}\quad\tau<0,  \\
\displaystyle \frac{1}{\lambda^{B}_{\beta}-\lambda^{B}_{\alpha}}
\displaystyle \sum_{\lambda^{B}_{\gamma}<0}
\,c^{B}_{\alpha \beta \gamma}\,\lambda^{B}_{\gamma}T
\,e^{\lambda^{B}_{\gamma}T\tau}
&\qquad {\rm for}\quad\tau>0.
\end{array}
\right.    \label{eq:conn}
\eeq

%
%p22
%
So comparing this formula Eq.(\ref{eq:conn})
with Eq.(\ref{eq:lam}) leads
to an inequality,
\beq
\bigl|\omega^{T}_{\alpha \beta}(\tau)\bigr|\,\ll\,
\bigl|T\,\lambda^{T}_{\alpha}(\tau)\bigr|\qquad \mbox{for large}\; T
\quad(\tau\neq0).
\eeq
This implies that the connection terms in Eq.(\ref{eq:act222}) can be
neglected for the large limit $T$. In other words, the eigenfunctions
$u^{T}_{\alpha}$ are covariant constants with respect to $\tau$ within
this approximation.

Returning to the original variable $t$, i.e., $\xi^{\alpha}(t)
=\sqrt{T}\xi^{\alpha}(\tau)$,\ $\lambda_{\alpha}(t)
=\lambda^{T}_{\alpha}(\tau)$
\ and\ $u_{\alpha}(t,x)=u^{T}_{\alpha}(\tau,x)$, we
get the approximate
action,
\beq
S^{(2)}\simeq \int_{-T}^{T}dt\,\sum_{\alpha}
\biggl\{\frac{1}{2}
\biggl(\frac{d\xi^{\alpha}}{dt}-\lambda_{\alpha}\,\xi^{\alpha}
\biggr)^{2}
+\psi^{\alpha}
\biggl(\frac{d\psi^{*\alpha}}{dt}-\lambda_{\alpha}\,\psi^{*\alpha}
\biggr)
\biggr\},
\eeq
and the path integral becomes
\beqa
K_{df}[\gamma_{A}\,\psi_{A},\,\gamma_{B}\,\psi^{*}_{B}]
=e^{-(E[\gamma_{B}]-E[\gamma_{A}])}\,\times\qquad\qquad  \nonumber  \\
\,\sum_{instanton}\int\,[d\xi][d\psi^{*}][d\psi]
\Bigl(
\sum_{\alpha}\frac{\partial f}{\partial\xi^{\alpha}}[\phi]\psi^{*\alpha}
\Bigr)
e^{-S^{(2)}}.  \label{eq:ker}
\eeqa

Before going into the concrete calculation of Eq.(\ref{eq:ker}), we make
some comments concerning the assumption (A-3) in section II.
Let us define the subsets of the indices
$\ii=\{\alpha=0,1,...\}$ in the eigenvalues $\lambda_{\alpha}(t)$,
\beq
\jj^{(+)}=\{\alpha\,|\,\lambda_{\alpha}(-t)>0\,\},\quad
\jj^{(-)}=\{\alpha\,|\,\lambda_{\alpha}(t)<0\,\} \quad \mbox{for}\quad t
\rightarrow \infty.
\eeq
Then, (A-3) means that the tangent spaces spanned by
$u_{\alpha}(t,\bullet)$
($\alpha \in \jj^{(+)}$) and
$u_{\alpha}(t,\bullet)$ ($\alpha \in \jj^{(-)}$)
satisfy the transversality
condition:
\beq
\ii=\jj^{(+)}\cup\jj^{(-)}
\eeq
and hence
\beq
\sharp\left(\jj^{(+)}\cap \jj^{(-)}\right)=1,\quad
\sharp\left(\ii\backslash \jj^{(+)}\right)=\ell.
\eeq
%
%p24
%
{}From now on we will write the indices as
\beqa
\ii^{(+)}&=&\ii \backslash \jj^{(-)}=\{\ell+1,\ell+2,...\},\\
\ii^{(-)}&=&\ii \backslash \jj^{(+)}=\{0,1,...,\ell-1\}, \\
\ii^{(+-)}&=&\jj^{(+)} \cap \jj^{(-)}=\{\ell\}.
\eeqa
Note that $\ii^{(+-)}$ is the set associated with the zero mode.

Now let us turn to the explicit calculation of the path integral
Eq.(\ref{eq:ker}). We start with the evaluation of the bosonic
fluctuations, which is the functional determinant of
the differential operator such as
\beq
\widehat{\cal O}_{\alpha}=
-\left(\frac{d}{dt}+\lambda_{\alpha}(t)
\right)
\left(\frac{d}{dt}-\lambda_{\alpha}(t)
\right)
\qquad{\rm for}\quad \alpha\in \ii.
\eeq
The determinant is easily calculated by the Gelfand-Yaglom
method.$^{{\rm 16,17}}$
Using the asymptotic behavior Eq.(\ref{eq:lam}), we get the formula
up to the $\alpha$-independent normalization\,:
%
%p26
%
\beq
\left(det\widehat{\cal O}_{\alpha} \right)^{-\frac{1}{2}}=
\left\{
\begin{array}{@{\,}l}
\sqrt{\lambda^{A}_{\alpha}}\,c_{\alpha}\,e^
{-\frac{T}{2}(\lambda^{A}_{\alpha}+\lambda^{B}_{\alpha})}
\qquad \quad {\rm for}\quad \alpha \in \ii^{(+)},   \\
\sqrt{|\lambda^{B}_{\alpha}|}\,c_{\alpha}\,e^
{\frac{T}{2}(\lambda^{A}_{\alpha}+\lambda^{B}_{\alpha})}
\qquad \quad {\rm for}\quad \alpha \in \ii^{(-)},   \\
\displaystyle \frac{1}{\sqrt{E[\gamma_{B}]-E[\gamma_{A}]}}\,
\sqrt{\lambda^{A}_{\alpha}|\lambda^{B}_{\alpha}|}\,c_{\alpha}\,e^
{-\frac{T}{2}(\lambda^{A}_{\alpha}-\lambda^{B}_{\alpha})}   \\
\qquad \qquad \qquad \qquad \qquad \quad \,\,
{\rm for}\quad \alpha \in \ii^{(+-)},
\end{array}
\right.
\eeq
where $c_{\alpha}$\,($\alpha\in \ii$) are some constants, which can be
eliminated by the contribution of the ghost fluctuations, and we have
also removed the zero mode from the determinant in the third formula
on the r.h.s..

Secondly, let us consider the contribution from the ghost fluctuations.
This may be written as
\beqa
\int \prod_{\alpha\in\ii}[d\psi^{*\alpha}]
[d\psi^{\alpha}]\,\psi^{*\ell}(t)\,
e^{-\int_{-T}^{T}dt\,\sum_{\alpha}\,\psi^{\alpha}
\bigl(\frac{d}{dt}\psi^{*\alpha}-\lambda_{\alpha}\,\psi^{*\alpha}
\bigr)}  \nonumber  \\
\quad \qquad  \qquad \qquad \qquad \mbox{with B.C. Eq.(\ref{eq:bc})}.
\label{eq:path}
\eeqa
It should be noticed that if there is no ghst insertion of
$\ii^{(+-)}$-component ($\ell$ index), the path integral vanishes
because of the zero mode as stated in the beginning of this section.
The calculation of Eq.(\ref{eq:path}) is
straightforward and we obtain
\beqa
\Bigl(\prod_{\alpha\in \ii}\,c^{-1}_{\alpha} \Bigr)\,
\int_{0}^{1}dx\,\Bigl(u_{\ell}(t,x),\,\delt(t,x) \Bigr)
\Bigl(\prod_{\alpha\in \ii^{(+)}}\,e^{\frac{T}{2}(\lambda^{A}_{\alpha}
+\lambda^{B}_{\alpha})}\Bigr) \qquad \qquad
\nonumber    \\
\times
e^{\frac{T}{2}(\lambda^{A}_{\ell}
-\lambda^{B}_{\ell})}
\Bigl(\prod_{\alpha\in \ii^{(-)}}\,e^{-\frac{T}{2}(\lambda^{A}_{\alpha}
+\lambda^{B}_{\alpha})}\Bigr)
\left(\psi^{*0}_{B}\cdots\psi^{*\ell}_{B}\right)
\left(\psi^{\ell-1}_{A}\cdots\psi^{0}_{A}\right).\quad
\eeqa
The boundary term
$\left(\psi^{*0}_{B}\cdots\psi^{*\ell}_{B}\right)
\left(\psi^{\ell-1}_{A}\cdots\psi^{0}_{A}\right)$
has the negative-eigenvalue indices,\ $\ii^{(-)}$ for $\psi_{A}$
and $\ii^{(-)}\cup\ii^{(+-)}$ for $\psi^{*}_{B}$, and so it may be
identified with (see Eq.(\ref{eq:orient}))
%
%p26
%
\beq
\omega^{(\ell+1)}[u^{B}_{\phi}]\otimes
\omega^{(\ell)}[u^{A}_{\phi}]^{*} \in\,
\Lambda^{\ell+1}T^{*}_{\gamma_{B}}\Omega_{p,q}(N)\otimes
\Lambda^{\ell}T_{\gamma_{A}}\Omega_{p,q}(N),
\eeq
where $u^{A}_{\phi}$ and $u^{B}_{\phi}$ denote the frames of
$\Gamma\left(\gamma^{*}_{A}TN\right)$ and
$\Gamma\left(\gamma^{*}_{B}TN\right)$ defined by the eigenfunctions
of $J_{\phi}$,\ $u_{\alpha}(t,x)\in T_{\phi(t,x)}N$\,:
\beqa
u^{A}_{\phi}&=&\lim_{t \to -\infty} (u_{0}(t,x),\,u_{1}(t,x),...)
=(u^{A}_{0}(x),\,u^{A}_{1}(x),...),  \\
u^{B}_{\phi}&=&\lim_{t \to \infty} (u_{0}(t,x),\,u_{1}(t,x),...)
=(u^{B}_{0}(x),\,u^{B}_{1}(x),...).
\eeqa
These frames may be approximated by parallel transporting the frame
$(u_{0},\,u_{1},...)$ along the instanton $\phi$ since each vector
$u_{\alpha}$ is almost covariant constant.

Putting everything together,
we have the heat kernel,
\beqa
K_{df}[\gamma_{A}\,\psi_{A},\,\gamma_{B}\,\psi^{*}_{B}]
&=&\frac{N}{\sqrt{2\pi}}\,
e^{-(E[\gamma_{B}]-E[\gamma_{A}]}\,(f[\gamma_{B}]
-f[\gamma_{A}])
\Bigl(
\prod_{\alpha\in\ii^{(+)}\cup\ii^{(+-)}}
\sqrt{\lambda^{A}_{\alpha}}\,
\Bigr)
\nonumber \\
& &
\Bigl(
\prod_{\alpha\in\ii^{(-)}\cup\ii^{(+-)}}\sqrt{|\lambda^{B}_{\alpha}|}\,
\Bigr)
\sum_{\phi}\omega^{(\ell+1)}[u^{B}_{\phi}]\otimes
\omega^{(\ell)}[u^{A}_{\phi}]^{*},
\eeqa
where $N$ is the normalization constant and $\phi$ runs over the set of
instantons connecting the two geodesics $\gamma_{A}$ and $\gamma_{B}$.
Recall that
\beq
\Omega_{\gamma_{I}}=\prod_{\alpha}\left(\frac{|\lambda^{I}_{\alpha}|}{\pi}
\right)^{\frac{1}{4}}
e^{-\frac{1}{2}|\lambda^{I}_{\alpha}|(\xi^{\alpha})^{2}}
\,\omega^{(m(I))}[\varepsilon^{I}]
\eeq
for the fixed frames of $\Gamma\left(\gamma_{I}^{*}TN\right)$,
\ $\varepsilon^{I}\,=\,(e^{I}_{0},\,e^{I}_{1},...)$
\ ($I=A,B\,;\,m(A)=\ell,\,m(B)=\ell+1$).
{}From Eq.(\ref{eq:mat}), after integrating $\Omega_{\gamma_{I}}$, we
finally obtain the matrix element,$^{{\rm 18}}$
%
%p27
%
\beqa
\langle \Omega_{\gamma_{B}}|\,\widehat Q \,|\Omega_{\gamma_{A}}\rangle
&=&-\frac{1}{\pi}\,e^{-(E[\gamma_{B}]-E[\gamma_{A}])}
\biggl(
\prod_{\alpha\in \ii^{(+)}\cup \ii^{(+-)}}
\lambda^{A}_{\alpha}
\biggr)^{\frac{1}{4}}
\biggl(
\prod_{\alpha\in \ii^{(-)}}|\lambda^{A}_{\alpha}|
\biggr)^{-\frac{1}{4}}
\nonumber  \\
& & {}\times
\biggl(
\prod_{\alpha\in \ii^{(-)}\cup \ii^{(+-)}}
|\lambda^{B}_{\alpha}|
\biggr)^{\frac{1}{4}}
\biggl(
\prod_{\alpha\in \ii^{(+)}} \lambda^{B}_{\alpha}
\biggr)^{-\frac{1}{4}}
\,\delta^{(\ell)}_{AB},
\label{eq:final}
\eeqa
where
\beq
\delta^{(\ell)}_{AB}=\sum_{\phi}\,
\langle \omega^{(\ell+1)}[\varepsilon^{B}],\,
\omega^{(\ell+1)}[u^{B}_{\phi}] \rangle
\langle \omega^{(\ell)}[u^{A}_{\phi}],\,
\omega^{(\ell)}[\varepsilon^{A}] \rangle
\eeq
with the product $\langle \,\bullet\,,\,\bullet\,\rangle$
of Eq.(\ref{eq:product}) ($G^{\alpha \beta}\simeq\delta^{\alpha \beta}$).

At this stage we make some remarks.
\begin{enumerate}
\item The terms other than $\delta^{(\ell)}_{AB}$ are factorized
into the part depending on the local data on $\gamma_{A}$
and that on $\gamma_{B}$, which may be absorbed in the normalizations
of $\Omega_{\gamma_{A}}$ and $\Omega_{\gamma_{B}}$, respectively.
\item There are no zero-eigenvalues in the Hessian operators evaluated
at $\gamma_{A}$ and $\gamma_{B}$. Therefore, Eq.(\ref{eq:final}) becomes
well defined using the $\zeta$-function regularization for the
positive eigenvalues.
\item $\delta^{(\ell)}_{AB}$ is interpreted as follows.
Since both $u^{A}_{\alpha}$\ and $e^{A}_{\alpha}$\,
($\alpha=0,1,...,\ell-1$)
are the orthonormal bases of the $\ell$-dimensional negative eigenspace
$N(\gamma_{A})\subset\Gamma(\gamma_{A}^{*}TN)$\,($u^{B}_{\alpha}$ and
$e^{B}_{\alpha}$ are in a similar situation), they are related by the
orthogonal matrix ${\cal O}^{A}_{\phi}\in O(\ell)$\,
(${\cal O}^{B}_{\phi}\in O(\ell+1)$). It follows that
\beq
\omega^{(\ell)}[\varepsilon^{A}]=\left(det{\cal O}^{A}_{\phi}\right)\,
\omega^{(\ell)}[u^{A}_{\phi}],\quad
\omega^{(\ell+1)}[\varepsilon^{B}]=\left(det{\cal O}^{B}_{\phi}\right)\,
\omega^{(\ell+1)}[u^{B}_{\phi}],
\eeq
and hence we have the expression,
\beq
\delta^{(\ell)}_{AB}=\sum_{\phi}\,\delta_{AB}(\phi),
\eeq
where $\delta_{AB}(\phi)=\left(det{\cal O}^{A}_{\phi}\right)
\left(det{\cal O}^{B}_{\phi}\right)$, which takes the values $\pm 1$
comparing the two orientations of $N(\gamma_{A})$ and those of
$N(\gamma_{B})$.
Following Witten's paper,$^{{\rm 3}}$
we now define a complex $(X,\,\delta)$\,:
\beqa
X&=&\bigl\{X^{\ell}\,|\,\ell=0,1,...\bigr\},   \\
\delta&=&\bigl\{\delta^{(\ell)}\,:\,X^{\ell}\,
\longrightarrow\,X^{\ell+1}\,|
\,\ell=0,1,...\bigr\}.
\eeqa
%
%p28
%
Here,\ $X^{\ell}$ is in Eq.(\ref{eq:hphys}) and
 $\delta^{(\ell)}$ the linear mapping
with matrix elements $\delta^{(\ell)}_{AB}$. Thus $\delta^{(\ell)}$ is
nilpotent,\ i.e.,\ $\delta^{(\ell+1)}\circ \delta^{(\ell)}=0$;\
we can form the cohomology associated with $(X,\,\delta)$.
\end{enumerate}
\section{EXAMPLE}
\setcounter{equation} {0}
Having formulated the complex $(X,\,\delta)$ on the mapping space
$\Omega_{p,q}(N)$ in general form, let us apply to a simple example.

We consider now $N$ to be the 2-dimensional sphere with the standard
metric, $S^{2}=\bigl\{(\phi^{1},\,\phi^{2},\,\phi^{3})\in \R^{3}\,|
\,(\phi^{1})^{2}+(\phi^{2})^{2}+(\phi^{3})^{2}=1\bigr\}$. First, remember
the basic facts concerning Morse theory over
$\Omega_{p,q}(S^{2})$.$^{{\rm 19}}$
\begin{enumerate}
\item[(M-1)]
Using the local coordinates $(\phi^{1},\,\phi^{2})$ defined by the
correspondence $(\phi^{1},\,\phi^{2},\,\phi^{3})\,\longrightarrow\,
(\phi^{1},\,\phi^{2})$, we express the metric on $S^{2}$ as
\beq
g_{ij}=\delta_{ij}+\frac{\phi^{i}\phi^{j}}
{1-(\phi^{1})^{2}-(\phi^{2})^{2}}.
\eeq
Then a Morse function over $\Omega_{p,q}(S^{2})$ is defined by
\beq
E[\phi]=\frac{1}{2}\int_{0}^{1}dx\,g_{ij}\del^{i}\del^{j},\qquad
\phi\in \Omega_{p,q}(S^{2}),
\eeq
where $p=(0,\,0)$ and $q=(\sin\kappa,\,0)$ ($0<\kappa<\pi$).
\item[(M-2)]
The geodesics in $\Omega_{p,q}(S^{2})$ are the paths which go round
the big circle $m$ times ($m$=0,1,...):
\beqa
\gamma^{(+)}_{m}(x)&=&(\sin(\kappa+2\pi m)x,\,0),  \\
\gamma^{(-)}_{m}(x)&=&(-\sin(\check{\kappa}+2\pi m)x,\,0)
\quad(\check{\kappa}=2\pi-\kappa)
\eeqa
with $\gamma^{(\pm)}_{m}(0)=p$ and $\gamma^{(\pm)}_{m}(1)=q.$
\item[(M-3)]
The Morse indices of $\gamma^{(+)}_{m}$ and $\gamma^{(-)}_{m}$ are $2m$
and $2m+1$, respectively. In fact we have the following eigenfunctions
of $J_{\gamma^{(\pm)}_{m}}$:$^{{\rm 20}}$
\beq
e_{n}(x)=\sin n\pi x \,(0,\,1)\qquad (n=1,2,...)
\eeq
with the corresponding eigenvalues,
\beqa
\lambda^{(+)}_{n}&=&(n\pi)^{2}-(\kappa+2m\pi)^{2}
\qquad {\rm for}\quad \gamma^{(+)}_{m},  \\
\lambda^{(-)}_{n}&=&(n\pi)^{2}-(\check \kappa+2m\pi)^{2}
\qquad {\rm for}\quad \gamma^{(-)}_{m}.
\eeqa
Note that the negative eigenvalues are $\lambda^{(+)}_{n}$ ($n=1,2,...,2m$)
and $\lambda^{(-)}_{n}$ ($n=1,2,...,2m+1$).
\end{enumerate}

Following the construction Eq.(\ref{eq:hphys}),
 we have the approximate physical
Hilbert space,
\beq
{\cal H}_{phys}\simeq \bigoplus_{\ell=0}^{\infty}X^{\ell},\quad
{\rm dim}X^{\ell}=1\quad {\rm for\ all}\ \ell.
\label{eq:phys}
\eeq
We now determine the coboundary operator $\delta^{(\ell)}:\,X^{\ell}
\,\longrightarrow\,X^{\ell+1}$. First, consider the 0-th coboundary
$\delta^{(0)}$. The relevant instantons are the solutions of
Eq.(\ref{eq:ins})\ ($\Gamma^{i}_{jk}=\phi^{i}\,g_{jk}$),
\beq
\delt^{i}=-\frac{\partial^{2}\phi^{i}}{\partial x^{2}}
-\phi^{i}\,g_{jk}\del^{j}\del^{k},
\label{eq:solution}
\eeq
satisfying the conditions $\phi(-\infty,\,x)=\gamma^{(+)}_{0}(x)$
and $\phi(\infty,\,x)=\gamma^{(-)}_{0}(x)$.
Then there exist two solutions
$\phi_{A}(t,x)$ ($A=R,L,$), which are opposite in sign of the second
component (see Fig.1):
\beq
\phi_{R}(t,x)=(\phi^{1}(t,x),\,\phi^{2}(t,x)),\quad
\phi_{L}(t,x)=(\phi^{1}(t,x),\,-\phi^{2}(t,x)),
\eeq
which have the asymptotic behavior,
\beq
\phi_{A}(t,x)\sim \left\{
\begin{array}{@{\,}ll}
{\rm exp}_{\gamma^{(+)}_{0}(x)}\left(c^{(+)}_{A}\,e^{\lambda^{(+)}_{1}t}
e_{1}(x)\right)\qquad & {\rm for}\quad t\rightarrow -\infty,  \\
{\rm exp}_{\gamma^{(-)}_{0}(x)}\left(c^{(-)}_{A}\,e^{\lambda^{(-)}_{1}t}
e_{1}(x)\right)\qquad & {\rm for}\quad t\rightarrow \infty,
\end{array}
\right.
\eeq
where $c^{(\pm)}_{A}$ are constants such that
 $c^{(\pm)}_{R}=-c^{(\pm)}_{L}$.
It is easy to see that these instantons
 induce the opposite orientations
on the 1-dimensional negative eigenspace
 $N\bigl(\gamma^{(-)}_{0}\bigr)$ and
therefore we have $\delta^{(0)}=0$.

Next, we shall examine the coboundary $\delta^{(2m)}:\,
X^{2m}\,\longrightarrow\,X^{2m+1}\,(m\ge 1)$. Then the instantons of
interest are the solutions of Eq.(\ref{eq:solution}), which asymptotically
satisfy the conditions
$\phi(-\infty,\,x)=\gamma^{(+)}_{m}(x)$
and $\phi(\infty,\,x)=\gamma^{(-)}_{m}(x)$.

These solutions are constructed as follows:
First slice the interval of $x$, $I=[0,\,1]$, into small pieces,
\beq
I_{i}=\biggl[\frac{i-1}{2m+1},\,\frac{i}{2m+1} \biggr],\quad
I=\bigcup_{i=1}^{2m+1}\,I_{i}.
\eeq
If we put
\beq
a_{i}=\gamma^{(+)}_{m}\left(\frac{i}{2m+1}\right)
=\gamma^{(-)}_{m}\left(\frac{i}{2m+1}\right)
\quad
(a_{0}=p,\,a_{2m+1}=q),
\eeq
then both of the restricted geodesics $\gamma^{(+)}_{m}\Big|_{I_{i}}$
and $\gamma^{(-)}_{m}\Big|_{I_{i}}$ express the paths which go from
$a_{i-1}$ to $a_{i}$ and further can be regarded as the geodesics
with the Morse indices 0 and 1, respectively. So there exist two
solutions of Eq.(\ref{eq:solution})
,\ $\phi^{I_{i}}_{A}\,(A=R,L)$, connecting
the geodesics $\gamma^{(\pm)}_{m}\Big|_{I_{i}}$.

Using these solutions as the building blocks we obtain two instantons
$\phi_{A}(t,x)\,(A=R,L)$ (see Figs.2 and 3):
\beq
\phi_{R}(t,x)=\left\{
\begin{array}{@{\,}ll}
\phi^{I_{1}}_{R}(t,x) \qquad & x\in I_{1}, \\
\phi^{I_{2}}_{L}(t,x)  & x\in I_{2}, \\
\phi^{I_{3}}_{R}(t,x)  & x\in I_{3}, \\
\qquad \vdots         &\quad \vdots     \\
\phi^{I_{2m+1}}_{R}(t,x) & x\in I_{2m+1}, \\
\end{array}
\right.
\eeq
and $\phi_{L}(t,x)$ is given by exchanging $\lq\lq$R" for $\lq\lq$L"
in the r.h.s..
Then they have the asymptotic behavior,
\beq
\phi_{A}(t,x)\sim \left\{
\begin{array}{@{\,}ll}
{\rm exp}_{\gamma^{(+)}_{m}(x)}\left(c^{(+)}_{A}\,e^{\lambda^{(+)}_{2m+1}t}
e_{2m+1}(x)\right)\qquad & {\rm for}\quad t\rightarrow -\infty,  \\
{\rm exp}_{\gamma^{(-)}_{m}(x)}\left(c^{(-)}_{A}\,e^{\lambda^{(-)}_{2m+1}t}
e_{2m+1}(x)\right)\qquad & {\rm for}\quad t\rightarrow \infty,
\end{array}
\right.
\eeq
where $c^{(\pm)}_{A}$ are constants such that
$c^{(\pm)}_{R}=-c^{(\pm)}_{L}$.

By the symmetry between $\phi_{R}$ and $\phi_{L}$, we get
$\delta^{(2m)}=0$. Again using similar analyses, it follows that
$\delta^{(2m+1)}=0$, so all coboundary operators vanish. Therefore the
result Eq.(\ref{eq:phys}) in a perturbative approximation is the exact
formula representing the physical Hilbert space. In fact we confirm
that this formula coincides with the de Rham cohomology
$H^{*}\left(\Omega_{p,q}(S^{2})\right)$.$^{{\rm 19}}$
\section{CONCLUDING REMARKS}
\setcounter{equation} {0}
In this paper we have taken a specific topological field theory over
$\Sigma=\R \times [0,1]$, whose physical Hilbert space is given
by the cohomology of the mapping space $\Omega_{p,q}(N)$.
Further work that is worth considering along these
lines is an examination
of the topological field theory over $\Sigma=\R \times M$ for a
Riemannian manifold $(M,\,h)$, where the action is defined by the
same formula as Eq.(\ref{eq:es0}).

This requires the consideration of Morse theory adapted to the mapping
space $\Omega(M,\,N)=\{\phi:(M,\,h)\longrightarrow
(N,\,g)\}$.$^{{\rm 21,22}}$
A natural choice of functional on $\Omega(M,\,N)$ is given by
\beq
E[\phi]=\frac{1}{2}\int_{M}dv_{h}\,g_{ij}(\phi)\,h^{\alpha \beta}(x)
\frac{\partial \phi^{i}}{\partial x^{\alpha}}
\frac{\partial \phi^{j}}{\partial x^{\beta}}.
\eeq
Then the critical points of $E$ are harmonic maps. It is known that
the Hessian operator has finite number of non-positive eigenvalues; one
can associate the Morse index for each harmonic map. Thus the main
problems are to construct $\lq\lq$ instantons" to interpolate between
the harmonic maps and to determine the coboundary operator of the
cohomology complex. This general area deserves further attention.

\

{\Large\bf ACKNOWLEDGMENTS}

S. Takahashi was partially supported by a Japan Ministry of Education,
Science and Culture Grant-in-Aid for Scientific Research on Priority
Areas (\#319), Project $\lq\lq$Symbiotic Biosphere: An Ecological
Interaction Network Promoting the Coexistence of Many Species".

Y. Yasui was partially supported by a Grant-in-Aid for Scientific
Research of the Ministry of Education, Science and Culture No. 04302015.
\bigskip

\appendix
\def\thesection {APPENDIX.}
\def\thesubsection {\arabic{subsection}}
\renewcommand{\theequation}{\mbox{\Alph{section}.\arabic{equation}}}
\setcounter{equation}{0}
\section{EXPANSION FORMULAS}
\label{Ap-C}
In this appendix we prove the expansion formulas Eqs.(\ref{eq:proof1}) and
(\ref{eq:proof2}).
Suppose $\phi_{s}$\,($s\in\,\R$) is a curve on $\Omega_{p,q}(N)$ such that
\beq
\phi_{s} = Exp_{\phi_{0}}\,(s\xi) = {\rm exp}\circ \sum_{\alpha=0}^{\infty}
s\xi^{\alpha}e_{\alpha},\qquad \xi \in \Gamma\left(\phi^{*}_{0}TN \right).
\eeq
Then, $\phi_{s}(x)$ ($s\in \R,\,x\in [0,1]$) is a family of curves on $N$
and a geodesic for any fixed $x$:
\beq
\nabla_{s}\frac{\partial\phi_{s}(x)}{\partial s}=0,\qquad
\frac{\partial\phi_{s}(x)}{\partial s}\Big|_{s=0} = \xi(x).
\eeq
We denote by $\partial/\partial\phi^{\alpha}_{s}\,
(\alpha=0,1,...)$ a basis of $\Gamma\left(\phi^{*}_{s}\,TN\right)$ and set
\beq
Y_{\alpha}(s,x)=s\frac{\partial}{\partial\phi^{\alpha}_{s}(x)}=
\frac{\partial}{\partial\lambda}
\left(exp\circ s\left(\xi+\lambda e_{\alpha}
\right)(x)\right)\Big|_{\lambda=0}. \label{eq:jaco}
\eeq
It is easy to show that each $Y_{\alpha}(s,x)$ for any fixed $x$ is
a Jacobi field on $N$ along $\phi_{s}(x)$:
\beq
\nabla_{s}\nabla_{s}\,Y_{\alpha}(s,x)+
R\left(Y_{\alpha}(s,x),\,\frac{\partial\phi_{s}(x)}{\partial s}\right)
\frac{\partial\phi_{s}(x)}{\partial s}=0,  \label{eq:y}
\eeq
with the initial conditions,
\beq
Y_{\alpha}(0,x)=0,\ \ \nabla_{s}\,Y_{\alpha}(0,x)=e_{\alpha}(x).
\label{eq:yy}
\eeq
So it is immediate that
\beq
\nabla^{2}_{s}\,Y_{\alpha}(0,x)=0,\qquad
\nabla^{3}_{s}\,Y_{\alpha}(0,x)
=-R\left(e_{\alpha}(x),\,\xi(x)\right)\xi(x).
\label{eq:yyy}
\eeq
\subsection {Expansion of {\protect \boldmath $G_{\alpha \beta}$}}
\begin{quotation}
We define $G_{\alpha \beta}(s)=
\langle \partial/\partial\phi^{\alpha}_{s},
\,\partial/\partial\phi^{\beta}_{s} \rangle$ and so, from
Eq.(\ref{eq:jaco}),
\beq
s^{2}\,G_{\alpha \beta}(s)
=\int_0^1dx\,\left(Y_{\alpha}(s,x),\,Y_{\beta}(s,x)
\right).  \label{eq:inner}
\eeq
The successive differentiations of Eq.(\ref{eq:inner}) yield:
\beqa
&6&\frac{d}{ds}\,G_{\alpha \beta}+
6s\left(\frac{d}{ds}\right)^{2}\,G_{\alpha \beta}
+s^{2}\left(\frac{d}{ds}\right)^{3}\,G_{\alpha \beta}  \nonumber \\
&=&\int_0^1dx\,\Bigl\{\left(\nabla^{3}_{s}\,Y_{\alpha},\,Y_{\beta}
\right)+3\left(\nabla^{2}_{s}\,Y_{\alpha},\,
\nabla_{s}\,Y_{\beta}\right) \nonumber \\
& &{}+
3\left(\nabla_{s}\,Y_{\alpha},\,\nabla^{2}_{s}\,Y_{\beta}
\right)+\left(Y_{\alpha},\,
\nabla^{3}_{s}\,Y_{\beta}\right)
\Bigr\},  \label{eq:del}
\eeqa
and
\beqa
&12&\left(\frac{d}{ds}\right)^{2}\,G_{\alpha \beta}+
8s\left(\frac{d}{ds}\right)^{3}\,G_{\alpha \beta}
+s^{2}\left(\frac{d}{ds}\right)^{4}\,G_{\alpha \beta}  \nonumber \\
&=&\int_0^1dx\,\Bigl\{\left(\nabla^{4}_{s}\,Y_{\alpha},\,Y_{\beta}
\right)+4\left(\nabla^{3}_{s}\,Y_{\alpha},\,
\nabla_{s}\,Y_{\beta}\right)
+6\left(\nabla^{2}_{s}\,Y^{\alpha},\,
\nabla^{2}_{s}\,Y_{\beta}\right)
\nonumber \\
& &{}+
4\left(\nabla_{s}\,Y_{\alpha},\,\nabla^{3}_{s}\,Y_{\beta}
\right)+\left(Y_{\alpha},\,
\nabla^{4}_{s}\,Y_{\beta}\right)
\Bigr\}. \label{eq:ddel}
\eeqa
It follows from these formulas, together with
Eqs.(\ref{eq:yy}) and (\ref{eq:yyy}),
that the expansion of $G_{\alpha \beta}(s)$ is given by
\beq
G_{\alpha \beta}(s) = \delta_{\alpha \beta}-\frac{s^{2}}{3}\int_0^1dx
\,\left(R(e_{\alpha},\,\xi)\xi,\,e_{\beta}\right)+{\cal O}(s^{3}).
\eeq
Setting $s=1$, we obtain Eq.(\ref{eq:proof1}).
\end{quotation}
%\subsection{Expansion of {\protect \boldmath $\grad_{\alpha}$}}
\subsection{Expansion of
 {\protect\boldmath ${\bf grad}\,E[\phi]_{\alpha}$}}
\begin{quotation}
Using Eqs.(\ref{eq:gg}) and (\ref{eq:jaco}), we have
\beq
s\left(\sgrad\,E[\phi_{s}]_{\alpha}\right)=\int^{1}_{0}dx\,\left(
-\nabla_{x}\frac{\partial\phi_{s}(x)}{\partial x}
,\,Y_{\alpha}(s,x)\right).
\label{eq:sgg}
\eeq
The successive differentiations of Eq.(\ref{eq:sgg}) yield:
\beqa
&\sgrad &\,E[\phi_{s}]_{\alpha}+s\frac{d}{ds}\sgrad\,E[\phi_{s}]_{\alpha}
\nonumber \\
&=&\int_0^1 dx\,
\biggl\{
\left(-\nabla_{x}\nabla_{x}\dels,\,Y_{\alpha}\right)
+
\left(R\Bigl(\delx,\,\dels \Bigr)\delx,\,Y_{\alpha} \right)
\nonumber \\
& & {}+
\left(-\nabla_{x}\delx,\,\nabla_{s}Y_{\alpha} \right)
\biggr\},
\eeqa
and
\beqa
&2&\frac{d}{ds}\sgrad\,E[\phi_{s}]_{\alpha}+s\left(\frac{d}{ds}\right)^{2}
\sgrad\,E[\phi_{s}]_{\alpha}
\nonumber \\
&=&\int_0^1 dx\,
\biggl\{
\left(-\nabla_{s}\nabla_{x}\nabla_{x}\dels,\,Y_{\alpha}\right)
+
2\left(-\nabla_{x}\nabla_{x}\dels,\,\nabla_{s}Y_{\alpha}\right)
\nonumber \\
& &{}+
\left(\nabla_{s}\left(R\Bigl(\delx,\,\dels \Bigr)\delx \right),\,
Y_{\alpha}\right)
+
2\left(R\Bigl(\delx,\,\dels \Bigr)\delx,\,\nabla_{s}Y_{\alpha}\right)
\nonumber \\
& &{}+
\left(-\nabla_{x}\delx,\,\nabla_{s}\nabla_{s}Y_{\alpha}\right)
\biggr\}.
\eeqa
Thus the expansion of $\sgrad\,E[\phi_{s}]$ is given by
\beqa
\sgrad\,E[\phi_{s}]_{\alpha}&=&\int_0^1 dx
\biggl\{
\left(-\nabla_{x}\frac{\partial\phi_{0}}{\partial x},\,e_{\alpha}\right)
-s\left(
\nabla_{x}\nabla_{x}\xi + R\Bigl(\xi,\,\frac{\partial\phi_{0}}{\partial x}
\Bigr)\frac{\partial\phi_{0}}{\partial x},\,e_{\alpha}
\right)
\biggr\}
\nonumber \\
& &{}+{\cal O}(s^{2}).
\eeqa
Setting $s=1$, we obtain Eq.(\ref{eq:proof2}).
\end{quotation}
{\large\bf REFERENCES AND FOOTNOTES}
\begin{enumerate}
\item E. Witten, Commun. Math. Phys. {\bf 117}, 353 (1988);
 {\bf 118}, 411 (1988).
\item For a review, see D. Birmingham, M. Blau, M. Rakowski, and
G. Thompson, Phys. Rep. {\bf 209}, 129 (1991).
\item E. Witten, J. Diff. Geom. {\bf 17}, 661 (1982).
\item P. Salomonson and J. W. Van Holten,
Nucl. Phys. B {\bf 196}, 509 (1982).
\item F. Cooper and B. Freedman, Ann. Phys. {\bf 146}, 262 (1983).
\item H. Bohr, E. Katznelson, and K. S. Narain, Nucl. Phys. B {\bf 238},
407 (1984).
\item J. M. Leinaas and K. Olaussen, Nucl. Phys. B {\bf 239},209 (1984).
\item T. Hirokane, M. Miyajima, and Y. Yasui, J. Math. Phys. {\bf 34},
2789 (1993).
\item B. Helffer and J. Sj\"ostrand, Commun. P.D.E. {\bf 10}, 245 (1985).
\item S. Gallot, D. Hulin, and J. Lafontaine, {\it Riemannian Geometry}
(Springer, New York, 1987).
\item The components of $R$ are defined by
$R\left(\frac{\partial}{\partial \phi^{i}},\,
\frac{\partial}{\partial \phi^{j}}\right)
\frac{\partial}{\partial \phi^{k}}
\,=\,R^{\ell}_{kij}\,\frac{\partial}{\partial \phi^{\ell}}$.
\item D. Birmingham, M. Rakowski, and G. Thompson,
 Nucl. Phys. B {\bf 315},
577 (1989).
\item I. A. Batalin and G. A. Vilkovisky,
 Phys. Rev. D {\bf 28},2567 (1983);
Nucl. Phys. B {\bf 234}, 106 (1984);
 J. Math. Phys. {\bf 26}, 172 (1985).
\item S. K. Ottarsson, J. Geom. Phys. {\bf 2}, 49 (1985).
\item Y. Chen and M. Struwe, Math. Z. {\bf 201}, 83 (1989).
\item I. M. Gelfand and A. M. Yaglom,
 J. Math. Phys. {\bf 1}, 48 (1960).
\item H. Kleinert, {\it Path Integrals}
 (World Scientific, Singapore, 1990).
\item We have used the normalization such that
$\,\langle \Omega_{\gamma_{I}}|\,
e^{-i\widehat H(2T)} \,|\Omega_{\gamma_{I}}\rangle\,=\,1$\ for\
$I=A,B.$
\item R. Bott, Amer. Math. Soc. {\bf 7}, 331 (1982).
\item There exist other eigenfunctions of
 $J_{\gamma^{(\pm)}_{m}}$ with the
positive eigenvalues: $e_{n}(x)=\sin n\pi x\,
\left(\displaystyle \frac{d}{dx}\gamma^{(\pm)}_{m}(x)\right)$ with
$\lambda^{(\pm)}_{n}=n^{2}\pi^{2}$ $(n=1,2,...)$.
\item J. Eells, {\it Harmonic Maps} (World Scientific, Singapore, 1992).
\item K. C. Chang, {\it Infinite Dimensional Morse Theory and Multiple
Solution Problems} (Birkh\"auser, Boston, 1993).
\end{enumerate}
\newpage
{\large \bf FIGURE CAPTIONS}
\begin{enumerate}
\item Numerical calculation of two instantons $\phi_{A}$ ($A=R,L$) going
from $\gamma^{(+)}_{0}$ to $\gamma^{(-)}_{0}$. The parameter $\kappa$ is
taken to be $\pi/2$: $p=(0,0,1)$ and $q=(1,0,0)$ on $S^{2}$. The dotted
lines of the instantons run the back side on $S^{2}$.
\item Numerical calculation of the instanton $\phi_{R}$ connecting
$\gamma^{(+)}_{2}$ with $\gamma^{(-)}_{2}$. The time flow is
$\gamma^{(+)}_{2}\rightarrow (a) \rightarrow (b) \rightarrow \cdots
\rightarrow (e) \rightarrow \gamma^{(-)}_{2}$.
\item Instanton configuration. The upper (lower) semicircles denote
the instantons $\phi_{R}$ ($\phi_{L}$) connecting the geodesics,
which are represented by $\lq\lq\bullet$". The numbers (0,1,...) denote
the Morse indices associated with the geodesics.
\end{enumerate}
\end{document}